\documentclass[copyright,creativecommons]{eptcs}

\usepackage[utf8]{inputenc}
\usepackage[T1]{fontenc}
\usepackage[scaled=0.8]{beramono}
\usepackage{amsmath}
\usepackage{amssymb}
\usepackage{stmaryrd}
\usepackage{mathtools}

\usepackage{booktabs}

\usepackage{underscore}
\usepackage{comment}

\def\thetitle{Deductive Verification via the Debug Adapter Protocol}

\usepackage[numbers,sort&compress]{natbib}
\usepackage{hyperref}
\hypersetup{hidelinks,
    colorlinks=true,
    allcolors=blue,
    pdfstartview=Fit,
    pdftitle=\thetitle,
    breaklinks=true}

\usepackage[capitalise]{cleveref}

\Crefname{equation}{Eq.}{Eqs.}
\Crefname{figure}{Fig.}{Figs.}
\Crefname{tabular}{Table}{Tabs.}
\Crefname{example}{Ex.}{Exs.}
\Crefname{section}{Sec.}{Sects.}
\Crefname{corollary}{Cor.}{Cors.}
\Crefname{proposition}{Prop.}{Props.}
\Crefname{theorem}{Thm.}{Thms.}
\Crefname{lemma}{Lem.}{Lems.}
\Crefname{algorithm}{Alg.}{Algs.}
\Crefname{definition}{Def.}{Defs.}

\providecommand{\event}{F-IDE 2021}

\title{\thetitle}
\author{Gidon Ernst \qquad Johannes Blau
\institute{Ludwig-Maximilians-Universität\\Munich, Germany}
\email{gidon.ernst@lmu.de}
\and
Toby Murray
\institute{University of Melbourne\\
Melbourne, Australia}
\email{toby.murray@unimelb.edu.au}
}
\def\titlerunning{\thetitle}
\def\authorrunning{G. Ernst, J. Blau \& T. Murray}
\begin{document}
\maketitle

\newcommand{\pipe}{\mathrel{\makebox[\widthof{$\Coloneqq$}]{$|$}}} % | with width of =

\newcommand{\C}[1]{\big(#1\big)}
\renewcommand{\phi}{\varphi}
\newcommand{\ps}{\underline{ps}}
\newcommand{\step}{\longrightarrow}
\newcommand{\xstep}[1]{\stackrel{#1}{\longrightarrow}}
\newcommand{\Step}{\mathit{step}}

\newcommand{\haslabel}{\mathbin{\dblcolon}}

\newcommand{\emptyseq}{\langle\rangle}
\newcommand{\singleseq}[1]{\langle#1\rangle}

\newcommand{\code}[1]{\texttt{#1}}
\newcommand{\loc}{\mathit{loc}}

\newcommand{\ASSUME}{\mathbf{assume}}
\newcommand{\ASSERT}{\mathbf{assert}}
\newcommand{\IF}{\mathbf{if}}
\newcommand{\THEN}{\mathbf{then}}
\newcommand{\ELSE}{\mathbf{else}}
\newcommand{\WHILE}{\mathbf{while}}
\newcommand{\DO}{\mathbf{do}}
\newcommand{\OR}{\mathbf{or}}
\newcommand{\SAT}{\mathbf{sat}}
\newcommand{\UNSAT}{\mathbf{unsat}}

\begin{abstract}
We propose a conceptual integration of deductive program verification
into existing user interfaces for software debugging.
This integration is well-represented in the ``Debug Adapter Protocol'',
a widely-used and generic technology to integrate debugging of programs into development environments.
Commands like step-forward and step-in are backed by
steps of a symbolic structural operational semantics,
and the different paths through a program are readily represented
by multiple running threads of the debug target inside the user interface.
Thus, existing IDEs can be leveraged for deductive verification debugging with relatively little effort.
We have implemented this scheme for SecC, an auto-active program verifier for~C, and discuss its integration into Visual Studio Code.
\end{abstract}

\section{Introduction}

Deductive verification of programs with respect to strong requirements
relies on human proof engineering effort.
The user has to provide the primary correctness specifications (e.g. procedure contracts),
as well as auxiliary annotations (e.g. loop invariants),
key lemmas, and other proof hints.
This is much facilitated by modern Integrated Development Environments (IDEs) for formal methods tools and by advances in verification technology.
Over the recent years, the term ``push-button'' has been coined,
suggesting perhaps that proof automation is nowadays good enough
to not burden the user with internal details.
However, proofs for code that is already correctly annotated are fundamentally different from the typical trial an error to find these.
The ability to dig into the causes of a verification failure is not just nice-to-have---it is crucial have access to as much information as possible.

\textbf{Related Work:} To that end, state-of-the-art IDEs for formal development offer different features:
Dafny~\cite{leino2014dafny}, for example, highlights those annotations which cannot be proven,
and the Boogie Verification Debugger~\cite{le2011boogie} gives structured access to concrete counterexamples.
In VeriFast~\cite{jacobs2011verifast}, one can inspect the symbolic state and a tree representation of the paths explored.
Why3~\cite{bobot2011why3} shows the generated verification conditions in a nice and structured way
and offers interactive as well as automatic proof steps.
In contrast, general purpose interactive theorem provers like Isabelle/PIDE~\cite{wenzel2018isabelle}, KIV~\cite{ernst:sttt2015}, Rodin~\cite{voisin2014rodin}, and PVS~\cite{masci2019pvs}
(to name a few with a sophisticated user interface),
tend to expose proof internals in detail.
The latter paper~\cite{masci2019pvs} nicely compares some popular formal IDEs and their features.

The approaches mentioned are rather different from the experience
of traditional, concrete debugging of programs in IDEs like Eclipse, IntelliJ, or Visual Studio Code, where the main features are breakpoints, single stepping, and inspection of data at runtime.
Recently, loose coupling between IDEs and language-specific toolchains has become popular,
based on the Language Server Protocol (LSP)%
    \footnote{\url{https://microsoft.github.io/language-server-protocol}}
and the Debug Adapter Protocol (DAP).%
    \footnote{\url{https://microsoft.github.io/debug-adapter-protocol}}
LSP is used in several of the above mentioned formal IDEs.
For VDM a debugger for concrete model executions has been developed~\cite{rask2021visual} using the DAP.
The KeY Symbolic Execution Debugger~\cite{hentschel2016interactive}
is a feature-rich tool for Java verification built on the Eclipse platform.

\textbf{Contribution:}
In this paper, we propose and describe an integration of
deductive program verification into general purpose IDEs on top of the DAP,
which can be used to interactively navigate to specific parts of a proof.
The first contribution is an abstract characterization of how that integration works conceptually (\cref{sec:model}).
As the second contribution, we briefly describe an ongoing implementation effort of this scheme for SecC,
an autoactive verifier for correctness and security of C~programs (\cref{sec:implementation})
inside Visual Studio Code.
Our conclusion is that implementing debugging
for an existing verification tool that is based on symbolic execution
is relatively easy and straight-forward when using the DAP.

\section{Conceptual Model of a Debug Server}
\label{sec:model}

\begin{figure}
    \centering
    \includegraphics[width=0.9\textwidth]{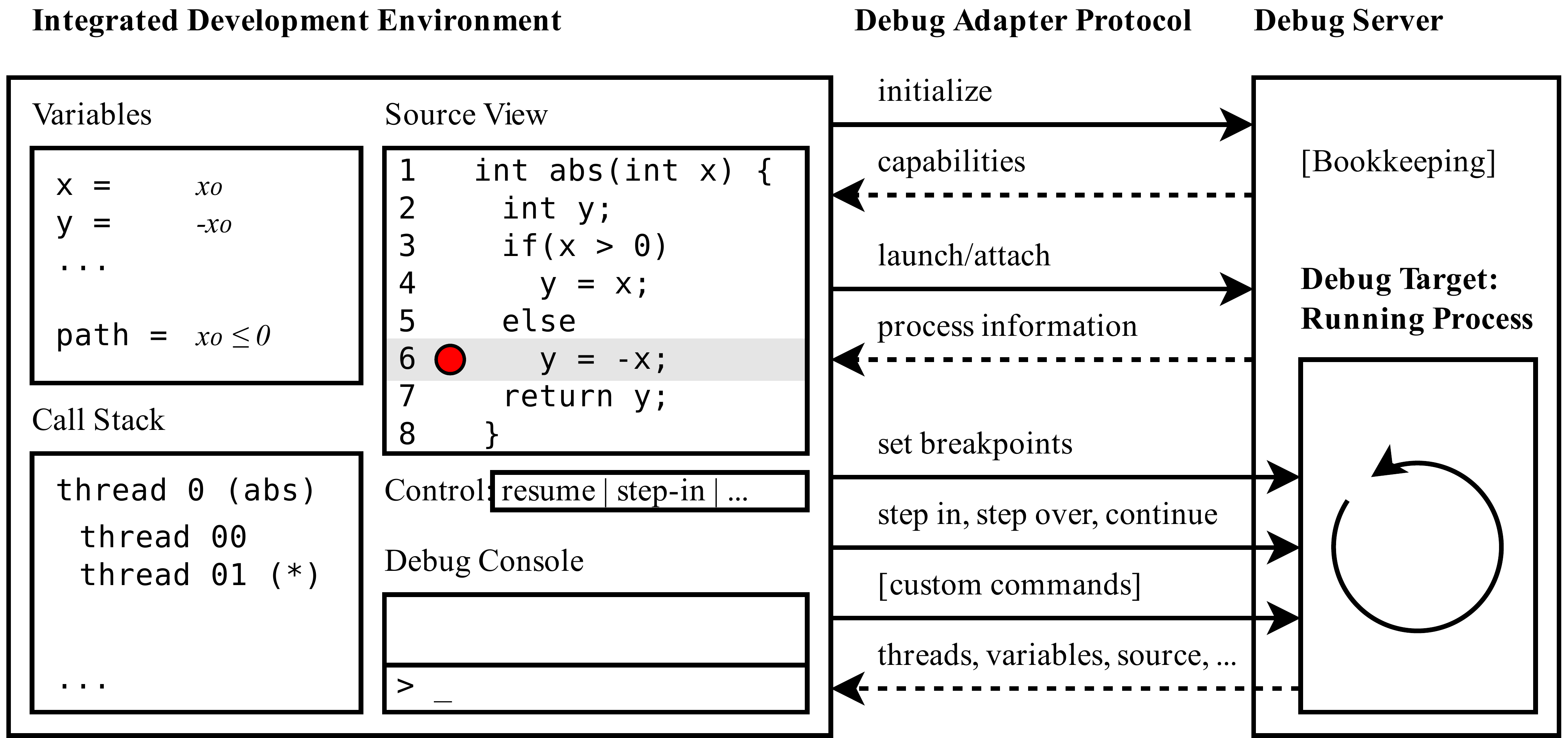}
    \caption{Schematics of the Debug View inside an IDE.}
    \label{fig:ide}
\end{figure}

In this section we develop a conceptual model of a debug server.
% i.e., a software component that provides debugging facilities via the Debug Adapter Protocol (DAP).
The server exposes certain operations to the IDE
and maintains the state of running processes that are being debugged (the debug targets).
\cref{fig:ide} depicts the integration between the IDE (as the client, on the left)
and a debug server (on the right) in terms of the graphical front-end of the IDE
and the messages exchanged between the two components.
Typically, the debugging perspective of an IDE shows the program's source code and the breakpoints within.
Moreover, there is a view that shows a current state of the program,
in terms of runtime values of variables (top-left), possibly organized according to the structure of lexical scopes.
% Below, the threads that are currently active are shown, which may be are running and or suspended.

In \cref{fig:ide} we anticipate a debug target that is executed symbolically, such that the program variables~\code{x} and~\code{y} are assigned symbolic expression over logical variables.
The example program, a function \code{abs} to compute the absolute value of parameter \code{x}, has been halted at a breakpoint in line 6, consequentially, variable \code{y} is set to $-x_0$ wrt. logical variable~$x_0$ capturing the initial value of \code{x}.
The synthetic variable \code{path} shows the current path condition,
a formula that captures the constraints for reaching the respective code location,
here the negated test of the conditional.
Below, the two possible branches through function \code{abs} are represented
as execution thread with identifiers~\code{00} and~\code{01}, that are subsumed by a parent thread~\code{0}.
The views in the window on the left are populated from data that is requested by the IDE from the server.

The simple, semi-formal model below omits inessential details of realistic languages (e.g., memory, see \cite{ernst:cav2019,berdine2005symbolic,jacobs2011verifast,fragoso2020gillian} here for details).
The key point is that the ideas presented here translate to any formal system
that can be described by a symbolic structural operational semantics~\cite{plotkin2004origins}.

\paragraph{Representation of States and Execution Steps.}

The state associated with a debug target primarily consists of a \emph{configuration}~$c$,
that describes the progress of the execution.
Configurations bundle up programs~$p$, defined over the set of program variables~$X$,
with \emph{symbolic assignment}~$s \colon X \to E$ from program variables to logical expressions~$E$
and a \emph{path condition}~$\phi$ as a logical formula that describes the branches taken so far.
Here, we capture execution using three kinds of configurations:
\begin{align*}
\text{Configurations } c
    & \Coloneqq
        \C{s,\phi,p_1; \ldots; p_k}
    \mid
        \C{c_1 \parallel \cdots \parallel c_n}
    \mid
        \C{\phi \land \lnot \psi,\lightning}
\end{align*}
\emph{Sequential compositions} execute programs~$p_1; \ldots; p_k$ from symbolic assignment~$s$ and path condition~$\phi$, where for $k=0$ this execution branch terminates.
\emph{Parallel compositions} of configurations capture multiple branches that arise e.g. from conditionals.
\emph{Proof obligations}, marked by~$\lightning$, verify that condition~$\psi$ holds for the path condition~$\phi$, where $\phi \land \lnot \psi\ \UNSAT$ denotes that the obligation discharges successfully.
In practice, such proof obligations would be annotated with additional contextual information.

Symbolic execution proceeds by unwinding a small-step relation
$c \xstep{\sigma} c'$ between configurations according to a \emph{schedule}~$\sigma$,
which reflects the user's choice where to step next.
Recursively, the first entry in the schedule, for $1 \le i \le n$,
resolves which part of a parallel configuration performs the next step:
\begin{align*}
\C{c_1 \parallel \cdots c_i \cdots \parallel c_n}
    & \xstep{\singleseq{i} \cdot \sigma}
        \C{c_1 \parallel \cdots c'_i \cdots \parallel c_n}
    \qquad \text{ if } \qquad c_i \xstep{\sigma} c'_i
\end{align*}
In the base case, execution of sequential composition steps with the empty schedule~$\sigma = \emptyseq$. Nondeterminism is captured by producing multiple branches, for example:
\begin{align}
\C{s,\phi,\ASSUME\ e; \ldots}
    & \xstep{\emptyseq} \C{s,\phi \land s(e), \ldots}
    \nonumber
    \\
\C{s,\phi,\ASSERT\ e; \ldots}
    & \xstep{\emptyseq} \C{s,\phi \land s(e),\ldots}
              \parallel \C{\phi \land \lnot s(e),\lightning}
    \nonumber
    \\
\C{s,\phi,x \coloneqq e; \ldots}
    & \xstep{\emptyseq} \C{s[x \mapsto s(e)],\phi,\ldots}
    \nonumber
    \\
\C{s,\phi,\IF\ e\ \THEN\ p_1\ \ELSE\ p_2; \ldots}
    & \xstep{\emptyseq} \C{s,\phi \land s(e),p_1; \ldots}
              \parallel \C{s,\phi \land \lnot s(e),p_2; \ldots}
    \nonumber
\end{align}
where $s(e)$ denotes evaluation of program expression~$e$ in symbolic state~$s$.
Of course, configurations with an unsatisfiable path condition
or empty sequential compositions can be soundly dropped
from their surrounding parallel context (which our implementation does eagerly),
e.g., the proof obligation from an $\ASSERT$ when $s(e)$ follows from $\phi$,
or either of the branches of an $\IF$ in case it is unreachable.
Loops can be unwound interactively (not discussed here) or summarized by invariants~$I$ as shown
\begin{align}
\C{s,\phi,\WHILE\ e\ \DO\ p; \ldots}
    & \xstep{\emptyseq} \C{s,\phi \land \lnot s(I); \lightning}
              \parallel \C{\hat s,\phi \land \hat s(e) \land \hat s(I),p; \ASSERT\ I}
              \parallel \C{\hat s,\phi \land \lnot \hat s(e) \land \hat s(I); \ldots}
    \nonumber
\end{align}
% (both approaches are e.g. supported in \cite{ernst:sttt2015,hentschel2016interactive}),
where the latter produces three successor configurations,
1) to prove invariant~$I$ initially,
2) to preserve $I$ over an arbitrary iteration,
where~$\hat s = s[x_1 \mapsto x'_1, \ldots, x_n \mapsto x'_n]$
introduces fresh logical variables $x_1', \ldots, x'_n$ for the program variables modified in loop body~$p$,
and 3) to continue with the code after the loop.

\paragraph{Initialization.}
The initial symbolic configurations for an entire translation unit is exemplified below.
A C~file has a list of global variables~$g_i$ that are to be initialized in sequence,
which we represent as a program~$g$ such that
$g \equiv \{ g_1 \coloneqq e_1; \ldots; g_n \coloneqq e_n \}$.
A top-level procedure declaration of the form $f(y_1, \ldots, y_m)~\{ q \}$,
with precondition/postcondition pair $P,Q$, parameters~$y_j$, and implementation~$q$
can be mapped onto a configuration~$c_f = \C{s,\mathit{true},\ASSUME\ P; q; \ASSERT\ Q}$
where $s_f = [g_i \mapsto x_i, y_j \mapsto y_j, \ldots]$ initializes all these variables in the respective scopes to fresh logical ones.
The verification of the \code{main} procedure may additionally assume
that the globals were just initialized,
which can be represented as $c_\code{main} = \C{s_\code{main},\mathit{true}, g; q}$
where the body $q$ of \code{main} is prefixed by the sequence~$g$ of global initializers.

\paragraph{Dispatching Requests in the Debug Server.}

We outline how to realize the operations that implement the main requests issued by the IDE in reference to a top-level configuration~$C$.
In addition, we track a set $B$ of breakpoints, which are program locations,
and we denote the location of a program by~$\loc(p)$.
In the following, we denote by $\sigma(C) = \C{s,\phi,\ldots}$ the sub-configuration
triggered by~$\sigma$, which is necessarily a sequential one in our simple model.

\begin{itemize}
\item
The request $\code{Launch}(\mathit{file})$ for program stored in~$\mathit{file}$ produces the initial configuration as a big parallel composition $C \coloneqq \C{c_{f_1} \parallel \cdots \parallel c_{f_n} \parallel c_\code{main}}$
where the respective sub-configurations are constructed wrt. the procedure declarations and globals as outlined above.

\item
The request \code{GetThreads} returns the currently running ``threads'',
which are those parts of parallel compositions that can be stepped.
This information can be represented in terms of possible schedules for the next step,
i.e., the set \code{GetThreads} returns $\{ \sigma \mid \exists\ c'.\ C \xstep{\sigma} c' \}$.
As suggested in \cref{fig:ide}, the hierarchical structure of these identifiers
can be exploited for more complex operations, such as stepping all branches
that share a common prefix schedule.

\item
The request $\code{GetVariables}(\sigma)$ returns the variable assignment~$s$
stored in the sequential configuration $\sigma(C)$ as remarked above.
In our implementation for SecC, we use this request to add synthetic variables,
such as $\code{path} \mapsto \phi$ where $\phi$ is the path constraint of
configuration~$\sigma(C)$ for that thread,
and we additionally include a representation
of the symbolic heap and information about the attacker level for security proofs.
% Moreover, in DSP, variables are actually organized into stack frames and scopes,
% which we discuss below.

\item
The request $\code{SetBreakpoints}(b)$ simply sets $B \coloneqq b$ to the set of specified breakpoints.
The protocol knows not only source breakpoints but also function breakpoints (triggered by calls), and exception breakpoints (when an exception is thrown),
which we have not used so far.

\item
The requests $\code{Next}(\sigma)$ and $\code{StepIn}(\sigma)$
execute a single transition
of a given thread~$\sigma$ according to the rules for $C \xstep{\sigma} c'$,
where~$c'$ is taken as the next state,~$C \coloneqq c'$.
The difference between these two commands in a concrete execution is that the first
proceeds over function calls in one atomic step,
whereas the second jumps into functions.
This behavior can be mirrored in a modular deductive verifier,
where \code{Next} dispatches such a call using a given function contract,
whereas \code{StepIn} inlines the call and disregards such contracts,
similarly for proving loops with invariants or unfolding a finite number of iterations
Our implementation supports \code{Next} only so far,
but e.g. KIV and KeY support both interactions in their respective GUIs.

\item
The request $\code{Continue}(\sigma)$ executes multiple transition
of a given thread~$\sigma$
until the corresponding configuration $\sigma(C) = \C{s,\phi,p;\ldots}$ with $\loc(p) \in B$,
i.e., the program execution has reached a breakpoint,
or until $\sigma(C) = \C{s,\phi}$ is final with no residual program.

\item
The request $\code{StepBack}(\sigma)$ undoes the latest corresponding transition (or sequence of transitions) of a particular sub-configuration.
This can be realized by keeping a history $c_0,c_1,\ldots$ of previous top-level configurations,
which in practice is facilitated by the fact that often tools are implemented
in functional languages and do not use destructive modification of states.

\item
The request $\code{Evaluate}(\sigma,e)$ is issued to inspect
the value of arbitrary expressions~$e$ within a state.
The response consists of evaluation $s(e)$ wrt. the logical variables
from $\sigma(C) = \C{s,\phi,\ldots}$.
Since the format of the result is just a string,
further information can be computed with the help of a solver and included in the response,
such as whether $\phi \implies s(e)$ holds when~$e$ is boolean,
or concrete values for~$e$ and its free variables if $\phi \land s(e)\ \SAT$.
\end{itemize}

\section{Debug Server Implementation for SecC}
\label{sec:implementation}

SecC%
    \footnote{\url{https://covern.org/secc}}
is an autoactive verifier programs for functional correctness
and security of C programs.
It is built around Security Concurrent Separation Logic (SecCSL) \cite{ernst:cav2019},
which can express value-dependent security properties of concurrent heap-manipulating programs.
The tool is currently used to verify a variety of small case studies.
Internally, SecC is based on a symbolic execution engine for Separation Logic~\cite{berdine2005symbolic}, which is similar to that of VeriFast (the latter is described nicely in~\cite{jacobs2011verifast}).
Thus, SecC lends itself to the approach outlined in \cref{sec:model}.

SecC is implemented in the Scala programming language,%
    \footnote{\url{https://bitbucket.org/covern/secc/}}
which runs on the Java Virtual Machine so that we can rely on the mature library \code{lsp4j},%
    \footnote{\url{https://github.com/eclipse/lsp4j}}
which fully abstracts the DAP (and also the Language Server Protocol, LSP)
in Java.
Creating a debug server with \code{lsp4j} simply amounts to implementing a particular Java interface whose operations correspond to DAP requests,
all protocol data structures are available as Java classes, too.
Within the client, here Visual Studio Code, some additional
effort has to be spent to register the respective language extension,
and to provide the necessary hooks that spin up the debug server with an appropriate configuration.

\begin{figure}
    \centering
    \includegraphics[width=\textwidth]{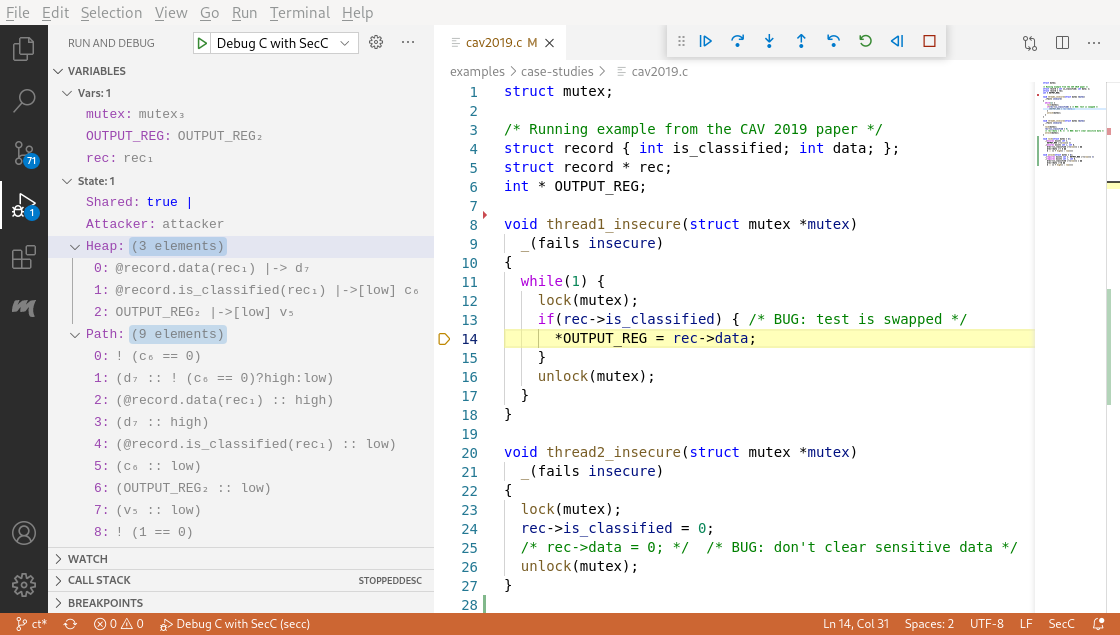}
    \caption{Debugging an insecure concurrent program in VSCode with SecC.}
    \label{fig:tree}
\end{figure}

The integration was developed as a VS Code extension in a Master thesis project by the second author over the course of roughly six months,
resulting in about 1000 LoC of Scala for the server and 400 LoC of TypeScript for the extension,
albeit getting a working initial version for a toy language was a matter of a few days.
In addition to debugging, the VS Code extension provides
syntax highlighting, verify-on-save, a debug console that lets one inspect the current state and evaluate expressions, and a graphical view of the symbolic execution tree.
Note that these extended features cannot be realized via the DAP alone but require LSP functionality as well as the extension facilities inside VS Code.
The extension is currently available in binary form (file \code{secc-0.2.0.vsix} on Bitbucket) and will be released as open source soon.

A screenshot of the debugging perspective is shown in \cref{fig:tree},
for an example from~\cite[Sec.~2]{ernst:cav2019} that has a defect.
The symbolic store~$s$ appears under ``Vars'' and the synthetic variables, including $\phi$ as ``Path'' under ``State'' on the left;
in addition there is a list of symbolic heap chunks describing the memory.
Stepping line 14 with the controls shown at the top-right
subsequently leads to a verification failure.
Informally, the code writes a secret value to a public memory location \code{OUTPUT_REG}.
This can be recognized from the data shown as follows:
Value~$d_7$ stored in \code{rec->data} (item~1 under Heap) is classified information (item 3: $d_7 \haslabel \code{high}$ under ``Path''),
whereas the memory location \code{OUTPUT_REG} is public (item~2: $\code{OUTPUT_REG}_2 \mapsto[\code{low}] \ldots$ under ``Heap'').

While we have not done a systematic evaluation or larger case study in this new SecC IDE yet,
the integration was useful for the second challenge of the VerifyThis 2021 competition~\cite{ernst2019verifythis}.
During the competition, it was helpful to investigate the symbolic states while developing the correctness proof interactively, for example to determine some subtle arithmetic constraints,
or to debug the unfolding/folding of memory predicates.

\section{Discussion \& Conclusion}

We have shown a concept to embed symbolic execution engines
into existing IDEs for interactive debugging via the established Debug Adapter Protocol (\cref{sec:model})
Our implementation for the autoactive verifier SecC proved to be straight-forward and low effort (\cref{sec:implementation}).

The approach inherits as a limitation the exponential path explosion from the nondeterministic execution when branches are not joined.
This is a problem with many conditionals in sequence, but we have not yet been impeded by this limitation.
At a conceptual level, it is not entirely clear how to remedy
the approach proposed here with ideas that defer splitting up branches
to the SMT solver as it is done in Boogie~\cite{barnett2005weakest} and generally in Horn clause verifiers~\cite{bjorner2015horn}.
Our approach can nevertheless complement such ideas,
e.g., by stepping selectively only that thread corresponding to a particular procedure or branch of interest to investigate precisely those proof obligations that fail with the more efficient encoding of~\cite{barnett2005weakest}.

The vanilla formulation of \cref{sec:model} repeatedly traverses the tree of configurations down to the currently executing leaf and rebuilds it on the way back.
Zippers~\cite{huet1997zipper} avoid this (probably perceived) inefficiency, and initial experiments with changing the implementation suggests that Zippers lead to quite elegant code, too.
This idea has indeed been followed before~\cite{ramsey2006applicative}.

Overall, we think that the proposed approach is general and flexible enough,
to be used to retrofit existing verification tools and languages with a symbolic interactive debugger.
By relying on existing infrastructure, such an undertaking is well within the reach of short-term projects.
By relying on established interaction paradigms, the approach brings software development practice and program verification a step closer together.
For future work we would like to investigate how to integrate concrete symbolic and concrete debugging techniques, and we plan to conduct a larger case study inside the SecC IDE to evaluate the benefits of the proposed approach in practice.

\smallskip
\noindent \emph{Acknowledgement.} We thank the reviewers for their suggestions to improve the presentation.

\bibliographystyle{eptcs}
\bibliography{references}

\end{document}